\documentclass[12pt]{iopart}

\usepackage{iopams}
\usepackage{epsfig}

\newcommand{\vr}{\bf r}
\newcommand{\ve}{\bf e}

\begin{document}

\title[Irreversible spherical model and entropy production rate]
{Irreversible spherical  model and its stationary entropy production rate}

\author{M. O. Hase$^a$ and M. J. de Oliveira$^b$}

\address{$^a$Escola de Artes, Ci\^encias e Humanidades,
Universidade de S\~ao Paulo, Avenida Arlindo B\'ettio 1000, 03828-000 
S\~{a}o Paulo, S\~{a}o Paulo, Brazil \\
$^b$Instituto de F\'{\i}sica,
Universidade de S\~{a}o Paulo,
Caixa Postal 66318,
05315-970 S\~{a}o Paulo, S\~{a}o Paulo, Brazil
}
\ead{mhase@usp.br, oliveira@if.usp.br}

\begin{abstract}
The nonequilibrium stationary state of an irreversible
spherical model is investigated on hypercubic lattices.
The model is defined by Langevin equations similar
to the reversible case, but with asymmetric transition rates. 
In spite of being irreversible, we have succeeded in finding 
an explicit
form for the stationary probability distribution, which
turns out to be of the Boltzmann-Gibbs type.
This enables one to evaluate the exact form of
the entropy production rate at
the stationary state, which is non-zero if the dynamical rules of the
transition rates are asymmetric.
\end{abstract}


\maketitle

\section{Introduction}

The spherical model was introduced by Kac \cite{K64} 
as a modification of the Ising model 
in which discrete spin variables are replaced by 
continuous ones, but subjected to the spherical constraint, which is a 
condition that ensures the thermodynamic properties of this system for any
temperature. The critical behaviour of this model was first analysed by 
Berlin and
Kac \cite{BK52}, and the exact solution can be found in any dimension $d$.
The model displays a continuous phase transition
for $d>2$, with non-classical critical behaviour for $2<d<4$
and mean field properties for $d>4$. 
The rich critical behaviour \cite{J72}, together with the 
establishment of many exact results, has made the spherical model 
a nice laboratory for statistical mechanics methods.

Being defined in a static way by the Boltzmann-Gibbs probability
distribution, the spherical model has no dynamics.
However, a dynamics can be assigned to the model
by the introduction of a set of Langevin equations \cite{R84},
which will rule the time evolution of the 
spin variables now transformed into stochastic variables.
The Langevin equations have additive white noise
and the deterministic parts are linear in the stochastic variables
\cite{R78,CS87}.
These variables, as in the static case, are associated to
the sites of a regular lattice and, in addition,
they are subject to the spherical constraint. 
The stationary probability distribution of the
associated Fokker-Planck equation turns out to be the Boltzmann-Gibbs 
probability distribution of spherical model.

The time dependent behaviour of the dynamic spherical model
defined by the Langevin equation has been examined in a series 
of papers \cite{CD95,GL00,ZKH00}.
In these works, the relaxation to the thermodynamic equilibrium
was investigated through two-point functions (autocorrelation 
and response function), which enables one 
to quantify a distance of the system 
from the equilibirum state \cite{CKP97}. 
This approach is based on 
an extension of the fluctuation-dissipation theorem to non-equilibrium 
states \cite{CR03,CG05}. 
We remark that the nonequilibrium situations analysed in these
papers, in which the system is relaxing to the equilibirum,
should be distinguished from the ones that concern us here, namely,
the situation in which the system
finds itself in a nonequilibrium stationary state.

The deterministic part of the Langevin equations, which
we call force, may be understood as
the gradient of the Hamiltonian defining the spherical model.
In other words, the force is conservative.
Since the Hamiltonian is a quadratic form, 
the force is linear in the stochastic variables
so that the linear coefficients make up a symmetric matrix.
In this paper, we consider Langevin equations for which 
the forces are still 
linear, but the coefficients lose the symmetric property 
becoming noconservative.
The set of Langevin equations with these nonconservative linear
forces, together with the spherical constraint, defines
the irreversible spherical model. We show here, that
in spite of the irreversibility, the stationary probability 
distribution can be written as being of the Boltzmann-Gibbs type. 
The description of the stationary distribution by a Boltzmann-Gibbs type 
function has already been found in models with Ising spin 
variables that lack detailed balance \cite{K84, LG06, GB09, G11, dO11}.

In the stationary state, the system is no longer
in the state of thermodynamic equilibrium 
because the forces are nonconservative. 
In this case, there will be a continuous production of entropy.
The second purpose of this paper is to calculate the production
of entropy in the stationary state, which, as we shall see,
can be done exactly. The entropy production rate for systems
described by a set of Langevin equations, or by the
associated Fokker-Planck equation, can be obtained from
an expression introduced by Tom\'e \cite{T06}, and also considered
by van den Broeck \cite{BE10}, which was derived
from an expression advanced by Schnakenberg \cite{S76} for
systems described by a master equation.
The critical behavior of the entropy production rate
is shown to be similar to that of the energy of the equilibrium
spherical model.

\section{Spherical model}

The spherical model \cite{BK52,J72} is defined as follows.
On a $d$-dimensional hypercubic lattice, with $N$ sites
and periodic boundary conditions,
a continuous spin variable $\sigma_{\vr}$ is attached to each site 
${\vr}$ of the lattice.
The usual nearest neighbour interaction Hamiltonian is written as
\begin{eqnarray}
{\cal H}(\sigma) = -\sum_{\vr}\sum_{\ve}J_{\ve}\sigma_{\vr}\sigma_{\bf r+e}
+\mu\sum_{\vr} \sigma^2_{\vr},
\label{H}
\end{eqnarray}
where the summation in $\ve$ is over the $d$ orthogonal unit vectors
that define the $d$-dimensional hypercubic lattice.
In a cubic lattice, for instance, these unit vectors are ${\ve}_1=(1,0,0)$,
${\ve}_2=(0,1,0)$ and ${\ve}_3=(0,0,1)$.  
We are considering the anisotropic case in which the
interactions are distinct for distinct directions.
The symbol $\sigma$ stands for the set of configurations 
$\{\sigma_{\vr}\}$ of the spins and
$J_{\ve}$ and $\mu$ are parameters. 

The probability distribution of configuration $\sigma$ is given by
\begin{equation}
P(\sigma) = \frac1Z e^{-\beta{\cal H}(\sigma)},
\label{4}
\end{equation}
where $\beta=1/k_BT$, $k_B$ being the
Boltzmann constant and $T$ the temperature.
The parameter $\mu$ is not free, but shoud be chosen such that 
\begin{eqnarray}
\sum_{r}\langle \sigma_{r}^{2}\rangle=N,
\label{meanconstraint}
\end{eqnarray}
which is called (mean) spherical constraint \cite{LW52,LW53}. 

The dynamics of the (mean) spherical model 
may be formulated through the Langevin equation
\begin{eqnarray}
\frac{d\sigma_{\vr}}{dt} = f_{\vr}(\sigma) + \eta_{\vr}(t),
\label{eq_langevin}
\end{eqnarray}
where the force $f_{\vr}(\sigma)$ is given by
\begin{eqnarray}
f_{\vr}(\sigma) = -\frac{\partial}{\partial\sigma_{\vr}}
{\cal H}(\sigma)
\end{eqnarray}
or
\begin{eqnarray}
f_{\vr}(\sigma)
= \sum_{\ve}J_{\ve}(\sigma_{\bf r + e}+\sigma_{\bf r - e}) - 2\mu\sigma_{\vr}.
\end{eqnarray}
As usual, the noise term $\eta_{\vr}(t)$ has the properties
\begin{eqnarray}
\langle\eta_{\vr}(t)\rangle = 0 
\qquad{\rm and}\qquad 
\langle\eta_{\vr}(t)\eta_{\vr'}(t')\rangle
= 2\Gamma\delta_{\vr,\vr'}\delta(t-t'),
\label{noise}
\end{eqnarray}
where $\Gamma=k_BT$ and $T$
is identified with the heat-bath temperature.

The time evolution of the probability $P(\sigma,t)$ 
of state $\sigma$ at time $t$ is given 
the Fokker-Planck equation
\begin{eqnarray}
\frac{\partial P(\sigma,t)}{\partial t}
= -\sum_{\vr}\frac{\partial}{\partial\sigma_{\vr}}
{\cal J}_{\vr}(\sigma,t),
\label{fp_eq}
\end{eqnarray}
where ${\cal J}_{\vr}(\sigma,t)$ is the probability current, given by
\begin{eqnarray}
{\cal J}_{\vr}(\sigma,t) = f_{\vr}(\sigma)
P(\sigma,t) - \Gamma\frac{\partial}{\partial\sigma_{\vr}}
P(\sigma,t).
\label{currenteq}
\end{eqnarray}
The probability distribution given by equation (\ref{4})
is the stationary solution of the Fokker-Planck equation.
In fact, in the present case, each probability current
at the stationary state,
\begin{eqnarray}
{\cal J}_{\vr}(\sigma) = f_{\vr}(\sigma)
P(\sigma) - \Gamma\frac{\partial}{\partial\sigma_{\vr}}P(\sigma),
\label{current_eq}
\end{eqnarray}
vanishes, and we may say that the system is in
thermodynamic equilibrium. 

\section{Irreversible spherical model}

In order to induce an irreversibility, and inspired by the Langevin 
equation (\ref{eq_langevin}), we introduce the irreversible dynamics by
\begin{eqnarray}
\frac{d\sigma_{r}}{dt} 
= f_{\vr}(\sigma) + \eta_{\vr}(t), 
\label{langevin}
\end{eqnarray}
where now the forces are given by
\begin{eqnarray}
\quad f_{\vr}(\sigma) 
= \sum_{\ve}(J_{\ve}\sigma_{\bf r+e}+J_{-\ve}\sigma_{\bf r-e})
- 2\mu\sigma_{\vr},
\label{force}
\end{eqnarray}
and cannot be anymore derived from a 
Hamiltonian unless $J_{\ve} = J_{-\ve}$ for all ${\ve}$. 
The parameters $J_{\ve}$ and $J_{-\ve}$, in this context, should 
be understood as the
strengths of the transition rates of the Markovian process defined
by the Langevin equation,
and not as an exchange integral entering the Hamiltonian
as in the reversible case.  
Notice that, as before, $\mu$ is not free but 
is a time dependent paramenter that should be chosen so that 
the constraint (\ref{meanconstraint}) is fulfilled.

The Fokker-Planck equation has the same form as before,
\begin{eqnarray}
\fl\frac{\partial P(\sigma,t)}{\partial t} 
= -\sum_{\vr}\frac{\partial}{\partial\sigma_{\vr}}
{\cal J}_{\vr}(\sigma,t) = -\sum_{\vr}\frac{\partial}{\partial\sigma_{\vr}} 
\left[ f_{\vr}(\sigma)
P(\sigma,t) - \Gamma\frac{\partial}{\partial\sigma_{\vr}}
P(\sigma,t)\right],
\label{fp}
\end{eqnarray}
but now the forces $f_{\vr}(\sigma)$ are nonconservative
and given by (\ref{force}).
In the stationary state, the probability current 
\begin{eqnarray}
\mathcal{J}_{\vr}(\sigma) = f_{\vr}(\sigma)
P(\sigma) - \Gamma\frac{\partial}{\partial\sigma_{\vr}}
P(\sigma)
\label{current_a}
\end{eqnarray}
does not vanish anymore, although the stationarity
condition,
\begin{eqnarray}
\sum_{\vr}\frac{\partial}{\partial\sigma_{\vr}} {\cal J}_{\vr}(\sigma) = 0,
\label{stat}
\end{eqnarray}
is fulfilled for the stationary probability distribution $P(\sigma)$.

The stationary probability distribution $P(\sigma)$
is obtained by assuming a form similar to (\ref{4}),
namely,
\begin{eqnarray}
P(\sigma) = Ce^{\Psi(\sigma)}
\qquad\textrm{with}\qquad \Psi(\sigma) 
= \sum_{\vr}\sum_{\ve}B_{\ve}\sigma_{\vr}\sigma_{\bf r+e} 
- A\sum_{\vr}\sigma_{\vr}^{2},
\label{pst}
\end{eqnarray}
where the summation is over the nearest neighbor pairs and
$A$ and $\{B_{\ve}\}$ are parameters to be found. 
We start by
writing the stationary Fokker-Planck equation (\ref{stat})
in the form
\begin{eqnarray}
\sum_{\vr} g_{\vr}(\sigma) = 0,
\label{station}
\end{eqnarray}
where
\begin{equation}
g_{\vr}(\sigma) = \frac{\partial f_{\vr}}{\partial\sigma_{\vr}} 
+ f_{\vr}\frac{\partial\Psi }{\partial\sigma_{\vr}}
- \Gamma\frac{\partial^2\Psi}{\partial\sigma_{\vr}^2}
- \Gamma\left(\frac{\partial\Psi }{\partial\sigma_{\vr}}\right)^2,
\label{g}
\end{equation}
which was
obtained after dividing the stationary equation (\ref{stat})
by $P(\sigma)$.
The substitution of $\Psi(\sigma)$, given by (\ref{pst}), 
and $f_{\bf r} (\sigma)$, given by (\ref{force}), into (\ref{g})
shows that $g(\sigma)$ is a quadratic form in the variable
$\sigma_{\bf r}$ plus a constant. This constant is $2(\mu-\Gamma A)$,
and should vanish. We conclude, therefore, that
\begin{equation}
A = \frac{\mu}{\Gamma}.
\label{A}
\end{equation}
Using this result, $g_{\vr}(\sigma)$ becomes the quadratic form
\begin{eqnarray}
\nonumber
\fl g_r(\sigma) &=& 
\sum_{\ve} B_{\ve}(c_{\ve}\,\sigma^2_{\bf r+e}+c_{-\ve}\,\sigma^2_{\bf r-e})
- 2A\sum_{\ve}\sigma_{\vr}(c_{\ve}\,\sigma_{\bf r+e}+c_{-\ve}\,\sigma_{\bf r-e}) \\ 
\fl &+& \sum_{\ve}B_{\ve}(c_{\ve}+c_{-\ve})\sigma_{\bf r-e}\sigma_{\bf r+e}
+ \sum_{{\ve,\ve' \atop \ve\perp\ve'}}B_{\ve}
(c_{\ve}\,\sigma_{\bf r+e}+c_{-\ve}\,\sigma_{\bf r-e})
(\sigma_{\bf r+e'}+\sigma_{\bf r-e'}),
\label{gg}
\end{eqnarray}
where 
\begin{equation}
c_{\ve} = J_{\ve}-\Gamma B_{\ve}
\qquad\textrm{and}\qquad
c_{-\ve} = J_{-\ve}-\Gamma B_{\ve}.
\end{equation}

The trivial solution of (\ref{station}) is obtained
by setting $c_{\ve}=0$ and $c_{-\ve}=0$, which
gives $J_{\ve}=\Gamma B_{\ve}=J_{-\ve}$, leading us back to the reversible model. 
To get a nontrivial solution, we substitute the expression 
(\ref{gg}) into (\ref{station}) and rewrite it in the form
\begin{eqnarray}
\nonumber
\fl & & 
\sum_{\vr}\sum_{\ve} B_{\ve}(c_{\ve}+c_{-\ve})\sigma^2_{\vr}
- 2A\sum_{\vr}
\sum_{\ve}(c_{\ve}+c_{-\ve})\sigma_{\vr}\sigma_{\bf r+e} \\ 
\fl &+& \sum_{\vr}\sum_{\ve}B_{\ve}(c_{\ve}+c_{-\ve})\sigma_{\bf r-e}\sigma_{\bf r+e}
+ \sum_{\vr}\sum_{{\ve,\ve' \atop \ve\perp\ve'}}B_{\ve}
(c_{\ve}+c_{-\ve})(\sigma_{\bf r+e}+\sigma_{\bf r-e})\sigma_{\bf r+e'} = 0,
\end{eqnarray}
which is solved by setting $c_{\ve}+c_{-\ve}=0$, leading to the condition
\begin{equation}
B_{\ve} = \frac{1}{2\Gamma}(J_{\ve} + J_{-\ve}).
\label{B}
\end{equation}

Therefore, the irreversible model defined by the
equations (\ref{langevin}) and (\ref{force}),
which embodies the parameters $J_{\ve}$, $J_{-\ve}$ and $\mu$,
has a stationary state of the Boltzmann-Gibbs type
given by (\ref{pst}) with 
the parameters $\{B_{\ve}\}$ and $A$ given by (\ref{A})
and (\ref{B}). It is worthwhile to notice that
a totally asymmetric dynamics is obtained by
setting $J_{-\ve}=0$, in which case $B_{\ve} = J_{\ve}/2\Gamma$, a result 
valid in any dimension.
The totally asymmetric dynamics has been shown to exist in systems with Ising 
spin variables in one and two dimensions \cite{K84, GB09}.

\section{Entropy production rate}

The variation of entropy $S$ of a system with time can
be splitted into two parts as
\begin{equation}
\frac{dS}{dt} = \Pi-\Phi,
\label{prig}
\end{equation}
where $\Pi$ is the entropy production rate and $\Phi$
is the entropy flux from the system to the environment.
Following \cite{T06}, the entropy production rate $\Pi$ of a 
non-equilibrium system governed by a Fokker-Planck equation can 
be evaluated by
\begin{equation}
\Pi(t) =  \frac{1}{\Gamma}\sum_{\vr} \int d\sigma
\frac{\left[{\cal J}_{\vr}(\sigma,t)\right]^{2}}{P(\sigma,t)}.
\label{pi}
\end{equation}
From the definition of entropy,
\begin{equation}
S(t) = - k_B\sum_\sigma P(\sigma,t)\ln P(\sigma,t),
\end{equation}
we get from (\ref{prig}) the following expression for
the entropy flux \cite{T06}:
\begin{equation}
\Phi(t) =  \frac{1}{\Gamma}\sum_{\vr} \int d\sigma
{\cal J}_{\vr}(\sigma,t)f_{\vr}(\sigma).
\label{phi}
\end{equation}
In the stationary state one has $\Pi=\Phi$, and we can use either
expression (\ref{pi}) or (\ref{phi}) to calculate
the entropy production rate.

From now on, we restrict ourselves to the simple case
in which $J_{\ve}$ and $J_{-\ve}$ are independent of 
${\ve}$, that is,
\begin{equation}
J_{\ve} = J \qquad\textrm{and}\qquad J_{-\ve}=J',
\end{equation}
but $J\neq J'$, and the parameter $B_{\ve}=B$ being independent 
of ${\ve}$, which leads to
\begin{equation}
B = \frac{J+J'}{2\Gamma}.
\end{equation}

\begin{figure}
\centering
\epsfig{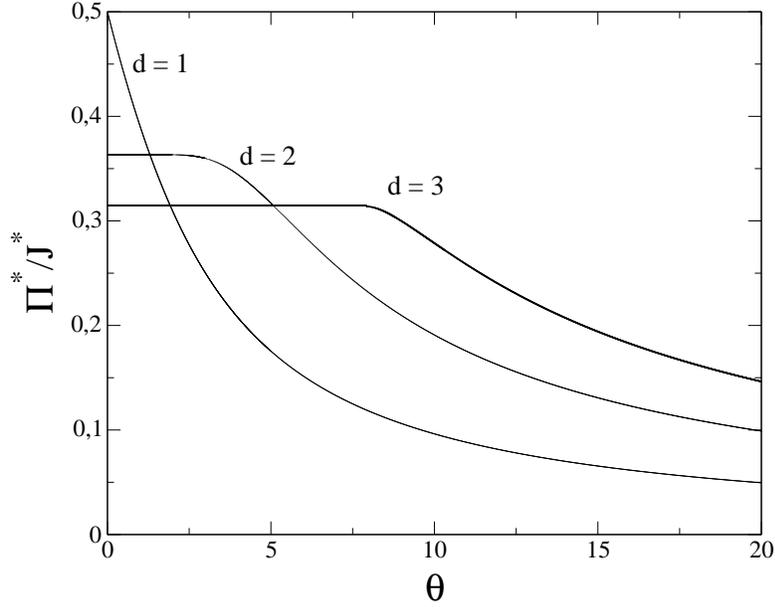}
\caption{
Plot of $\Pi^*/J^*$
as a function of $\theta$
in $d=1$, $d=2$ and $d=3$.}
\label{figR}
\end{figure}

Using the results of the previous section in the expression
(\ref{pi}), the entropy production rate per site $\Pi^*=\Pi/N$ can 
be evaluated as
\begin{equation}
\Pi^* = \frac{d}{2\Gamma}
\left(q_{0}-q_{2}\right)(J-J')^2,
\end{equation}
where $q_0$ and $q_2$ are defined by
\begin{equation}
q_0 = \langle\sigma^2_{\vr}\rangle   \qquad\textrm{and}\qquad 
q_2 = \langle\sigma_{\bf r-e}\sigma_{\bf r+e}\rangle,
\end{equation}
with ${\ve}$ being any one of the unit vectors.
Notice that $\Pi^*$ vanishes when the reversibility
condition $J=J'$ is satisfied, as expected.

Using the stationary probability distribution (\ref{pst}),
we get the following results for $q_0$ and $q_2$:
\begin{eqnarray}
q_{0} = 
\frac{1}{4B}\int_{0}^{\infty}d\xi\,e^{-\frac{A}{2B}\xi}
\left[I_{0}(\xi)\right]^{d}
\label{q0}
\end{eqnarray}
and
\begin{eqnarray}
q_{2} = 
\frac{1}{4B}\int_{0}^{\infty}d\xi\,e^{-\frac{A}{2B}\xi}
\left[I_{0}(\xi)\right]^{d-1}I_{2}(\xi),
\label{q2}
\end{eqnarray}
where the modified Bessel function of first kind of 
order $n$ is denoted by $I_{n}$. 
The parameter $\mu$ should ensure the spherical constraint, 
which is simply $\langle\sigma_{\vr}^2\rangle=1$ or $q_0=1$. 
Setting the right-hand side of equation (\ref{q0}) equal
to $1$, one has $B$ as an implicit function of $A$.
If we define $\theta=1/B=2\Gamma/(J+J')$, the 
production of entropy per site becomes 
\begin{equation}
\Pi^* = \frac{d}{\theta}\left(1-q_{2}\right)J^*,
\end{equation}
where
\begin{equation}
J^* = \frac{(J-J')^2}{J+J'}.
\end{equation}

Note that the quantity $\Pi^{\ast}/J^{\ast}$ is a function of $\theta$ only, 
and a graph $\Pi^{\ast}/J^{\ast}\times\theta$ is plotted in 
Figure \ref{figR} for dimensions $d=1,2,3$.
In one dimension, an analytical expression is available, and is given by
\begin{equation}
\frac{\Pi^*}{J^*} = \frac18\left(\sqrt{16+\theta^2}-\theta\right).
\end{equation}
When $\theta\to0$, the quantity $\Pi^{\ast}/J^{\ast}$ aproaches a
constant $c_d$ (in $d$ dimensions), which is given by
\begin{eqnarray}
c_{d}=\left\{
\begin{array}{lcl}
\displaystyle\frac{1}{2} & , & d = 1 \\
 & & \\
1-\displaystyle\frac{2}{\pi} & , & d = 2 \\
 & & \\
\displaystyle\frac{d}{4} \int_{BZ}\frac{1-\cos2k_1}{d-\sum_{i=1}^d\cos k_i}
\frac{d^dk}{(2\pi)^d} & , & d\geq 3
\end{array}
\right.,
\end{eqnarray}
where the integration above is over the Brillouin zone ($BZ$).
For $d=3$, we get $c_3=0.314762\cdots$.
Notice that in $d\geq3$, the quantity $\Pi^*/J^*$ equals the 
constant $c_d$ in the ferromagnetic phase, $\theta\leq\theta_c$,
where $\theta_c$ is given by
\begin{equation}
\frac1{\theta_c} = \frac{1}{4} \int_{BZ}\frac{1}{d-\sum_{i=1}^d\cos k_i}
\frac{d^dk}{(2\pi)^d}.
\end{equation}
For $d=3$, we get $\theta_c=7.913552\cdots$.

\section{Conclusion}

In this paper, we have investigated a system in a nonequilibrium
stationary state. The (mean) spherical model is a suitable
laboratory where many exact results are available, and we have
succeeded in finding an exact form for the
probability distribution, despite the fact that 
the system is not in equilibrium
state (as testified by the non-zero value of the entropy
production). It is worthwhile 
to mention that the probability distribution found is of the
Boltzmann-Gibbs type. The knowledge of this particular form 
allowed us to explicitly evaluate the stationary entropy production.
The origin
of the nonequilibrium behaviour in our work, which is responsible for
the non-zero entropy production, goes back to the unbalanced
transition rate to the opposite direction,
$J_{\ve}\neq J_{-\ve}$ (for any $\ve$). If, on the other hand, the
condition $J_{\ve}=J_{-\ve}$ is satisfied for any $\ve$, the stationary 
entropy production vanishes.

\end{document}